\documentclass[conference]{IEEEtran}

\usepackage{graphicx}
\usepackage{multirow}
\usepackage{float}
\usepackage{caption}
\usepackage{subcaption}
\usepackage{tikz}
\usepackage{pgfplots}
\usepackage{epsfig}
\usepackage[percent]{overpic}
\usepackage{amsmath}
\usepackage{amsfonts}
\usepackage{algorithmic}
\usepackage{algorithm}
\usepackage{hyperref} 
\usepackage[english]{babel}

\begin{document}
	
\title{Epitaxial growth of BaBiO$_3$ thin films on SrTiO$_3$(001) and MgO(001) substrates using molecular beam epitaxy and controlling their crystal orientations competition}
	
\author{\IEEEauthorblockN{I. Ahmed$^{\text{(a),(b)*}}$, S. De Gendt$^{\text{(b),(c)*}}$, C. Merckling$^{\text{(a),(b)*}}$}
\IEEEauthorblockA{(a) Department of Materials Engineering, KU Leuven, Kasteelpark Arenberg 44, 3001 Leuven, Belgium. \\
(b) Imec, Kapeldreef 75, 3001 Leuven, Belgium. \\
(c) Department of Chemistry, KU Leuven, Celestijnenlaan 200F, 3001 Leuven, Belgium. \\
Corresponding author*: islam.ahmed@imec.be}}
	
\maketitle
	
\begin{abstract}

 BaBiO$_3$ has lately gained high research attention as a parent material for an interesting family of alloyed compositions with multiple technological applications. In order to grow a variety of structures, a versatile deposition tool such as molecular beam epitaxy has to be employed.  In this work, molecular beam epitaxy growth of BaBiO$_3$ on SrTiO$_3$(001) and MgO(001) substrates is studied. When grown by molecular beam epitaxy on SrTiO$_3$(001) or MgO(001) substrates, BaBiO$_3$ is known to have two competing orientations namely, (001) and (011). Characterization of the thin film is carried out by X-ray diffraction, X-ray reflectivity, atomic force microscopy, Rutherford backscattering, and transmission electron microscopy. Pathways to block the growth of BaBiO$_3$(011) and only grow the technologically relevant BaBiO$_3$(001) are described for both substrates. Understanding of the enabler mechanism of the co-growth is established from epitaxy point of view. This can be beneficially utilized for growth of the different compositions of BaBiO$_3$ material family in a more controlled manner.

\end{abstract}

\IEEEpeerreviewmaketitle

\section{Introduction}

 Thanks to their chemical and structural versatility, complex oxides have attracted great research interest over the past few decades \cite{vila2015integration, zubko2011interface}. Perovskite oxide is a family of compounds belonging to the class of complex oxides with an ABO$_3$ crystal structure \cite{pena2001chemical}. BaBiO$_3$ (BBO) is a perovskite compound which was first synthesized in 1963 \cite{scholder1963alkali}. BBO has a monoclinic structure at room temperature and a cubic structure at temperatures above 820 K \cite{cox1979mixed}. In contradiction with band theory of solids, BBO has a band gap of 2 -- 2.19 eV \cite{vesto2022observation, ahmed2024influence}. Thanks to its electronic properties, BBO has been investigated for wide range of technological applications like efficient photocatalytic harvesters \cite{tang2007efficient}, photovoltaic devices \cite{chouhan2018babio3}, water splitting\cite{huerta2019visible}, and CO$_2$ degradation \cite{khraisheh2015visible}. BBO shows ferroelectric behavior when grown on Pt-coated Si substrate \cite{acero2024unveiling}, while BBO nanoparticles exhibit ferromagnetic response \cite{shilna2022novel}, which is promising for memory applications. BBO might also serve as a favorable buffer layer for other perovskite oxide thin films with large lattice constants, such as YBiO$_3$ (\textit{a} = 4.4 \AA) \cite{bouwmeester2019stabilization}. In a computational based study, it was predicted that a 2-dimensional electron gas (2DEG) is established for BiO$_2$ terminated BBO(001), via self-doping, which is advantageous for oxitronic devices \cite{vildosola2013mechanism}.

 Besides self-doping, alloying could potentially widen the application area of a certain parent material by strongly tailoring its electronic, optical, or structural properties \cite{chambers2010epitaxial}. Perovskite oxides serve as versatile parent compounds thanks to the possible substitution on either of the three different ionic sites which exists within the ABO$_{3}$ structure. If this is applied to our parent material of interest, BBO, many examples could be found in literature. One example is the discovery of superconductivity for n-type doped BaBi$_{1-x}$Pb$_x$O$_3$, with a critical temperature of 13 K \cite{sleight1975high}. Ba$_{1-x}$K$_x$BiO$_3$ Ba$_{1-x}$Rb$_x$BiO$_3$ (p-type doped BBO) were then later discovered as superconductor oxides with a transition temperature of 29.8 K \cite{cava1988superconductivity}. Partial n-type anionic substitution led to the discovery of topological insulating BaBi(O$_{1-x}$F$_{x}$)$_3$, according to a density functional theory (DFT) study \cite{yan2013large}. If experimentally realized, BaBi(O$_{1-x}$F$_{x}$)$_3$ could feasibly be implemented in technologies such as magnetics, thermoelectrics, or quantum computing \cite{tian2017property, xu2017topological, hasan2010colloquium}. In addition, if complete \textit{ex situ} topotactic fluorination was carried out for BBO thin film, fluorite structure BaBiF$_5$ could be obtained, which can be used as a fluorine ionic conductor in ion based batteries \cite{chikamatsu2018fabrication}. Visibly transparent p-type semiconducting layers based on Ba$_{1-x}$K$_x$Bi$_{1-x}$Ta$_x$O$_3$ with maximum hole mobility of around 30 cm$^{2}$/V.s and increased band gap of 4.5 eV \cite{bhatia2016high} was discovered. Along side with its counterpart n-type transparent In-Ga-Zn-O (with electron mobility higher than 100 cm$^{2}$/V.s) \cite{nomura2004room}, this material system is very important for enabling complementary metal-oxide-semiconductor (CMOS) based thin film transistor technology. Molecular beam epitaxy (MBE) has the ability to grow an alloyed thin film with a variety of compositions in a controlled manner \cite{nunn2021review}.

 Understanding the growth mechanism, phase stability, and kinetics of the parent perovskite thin film is an essential prerequisite for developing high quality alloyed structures. Based on its co-deposition capability at ultra high vacuum environment, MBE is considered advantageous in controlling the stoichiometry and crystal orientation (important aspects for realizing the desired technological edge) of the grown thin film \cite{nunn2021review}. Accessing a growth window where the cations' stochiometry is self-regulating has been made possible by MBE thanks to the presence of a volatile elemental component \cite{brahlek2018frontiers}. In our previous study, epitaxy of BBO was shown to follow adsorption-controlled regime when grown on SrTiO$_3$(STO)-buffered Si(001) substrates \cite{ahmed2022self}. Alongside with MBE parameters that play crucial role in the epitaxy, substrate choice is also very important as it has influence on the crystal orientation and film quality. BBO(001) \& BBO(011) are known to be competing orientations in the thin film growth, based on previous studies \cite{hellman1989molecular, iyori1995preparation, makita1997control, mijatovic2002growth, muta2022growth, ahmed2022self}, depending on the substrate of choice. In this study, MBE growth of BBO thin films on both STO ($a_\text{STO}$ = 3.905 \AA) and MgO ($a_\text{MgO}$ = 4.212 \AA) substrates is investigated. Crystal quality and phase formation for the various BBO thin films grown using different growth parameters are assessed. The influence of using a BaO buffer layer on the crystal orientation and morphology of the thin film grown on STO substrate is also demonstrated.




\section{Experimental}

 To understand the growth mechanism for BBO thin films on both STO and MgO bulk substrates (Furuuchi chemical Co.), ultra-high vacuum MBE reactor was utilized. 10 mm $\times$ 10 mm Bulk substrates were cleaned before introduction into the tool by dipping them into isopropanol for few hours followed by an overnight acetone bath. Substrate coupons were then glued on a carrier wafer. Before MBE's growth starts, surface of the glued coupons are further cleaned by heat treatment in oxygen environment at 800$^{\circ}$C for 30 minutes. Metallic sources were evaporated out of their thermal effusion cells towards the rotating substrates after calibrating the fluxes using an \textit{in situ} quartz crystal microbalance (QCM) setup. Growth was carried out in an activated oxygen species environment, generated using a remote radio-frequency (RF) plasma cell, with a background pressure of 3E-6 Torr and plasma power of 600 W. Thin films were cooled down from growth temperature (T$_s$) to room temperature at a rate of approximately 10$^{\circ}$C/minute at the same oxygen conditions.

 The crystalline structure and quality of the thin films was checked utilizing X-ray diffraction (XRD) setup with Cu-K$_\alpha$ radiation. Symmetric $\omega$–2$\theta$ scans were collected between 13$^{\circ}$ and 31$^{\circ}$ scattering angles for all samples, alongside with the rocking curves (RC's) around the interesting diffraction peaks. Using the same setup, film thickness and roughness of the thin films were deduced based on X-ray reflection (XRR) measurements. Transmission electron microscopy (TEM) technique with a 200 kV operating voltage was used to further evaluate the crystal structure for one of the thin films. Sample preparation for TEM inspection was minimized to avoid any damage to the e-beam sensitive BBO layer. Surface morphology investigation was carried out with atomic force microscope (AFM) operating in tapping mode. Evaluation of the elemental composition of the grown thin films was done by Rutherford backscattering (RBS) method.

\section{Results}

\begin{figure}[h]      
	
	\begin{overpic}[width=0.245\textwidth]{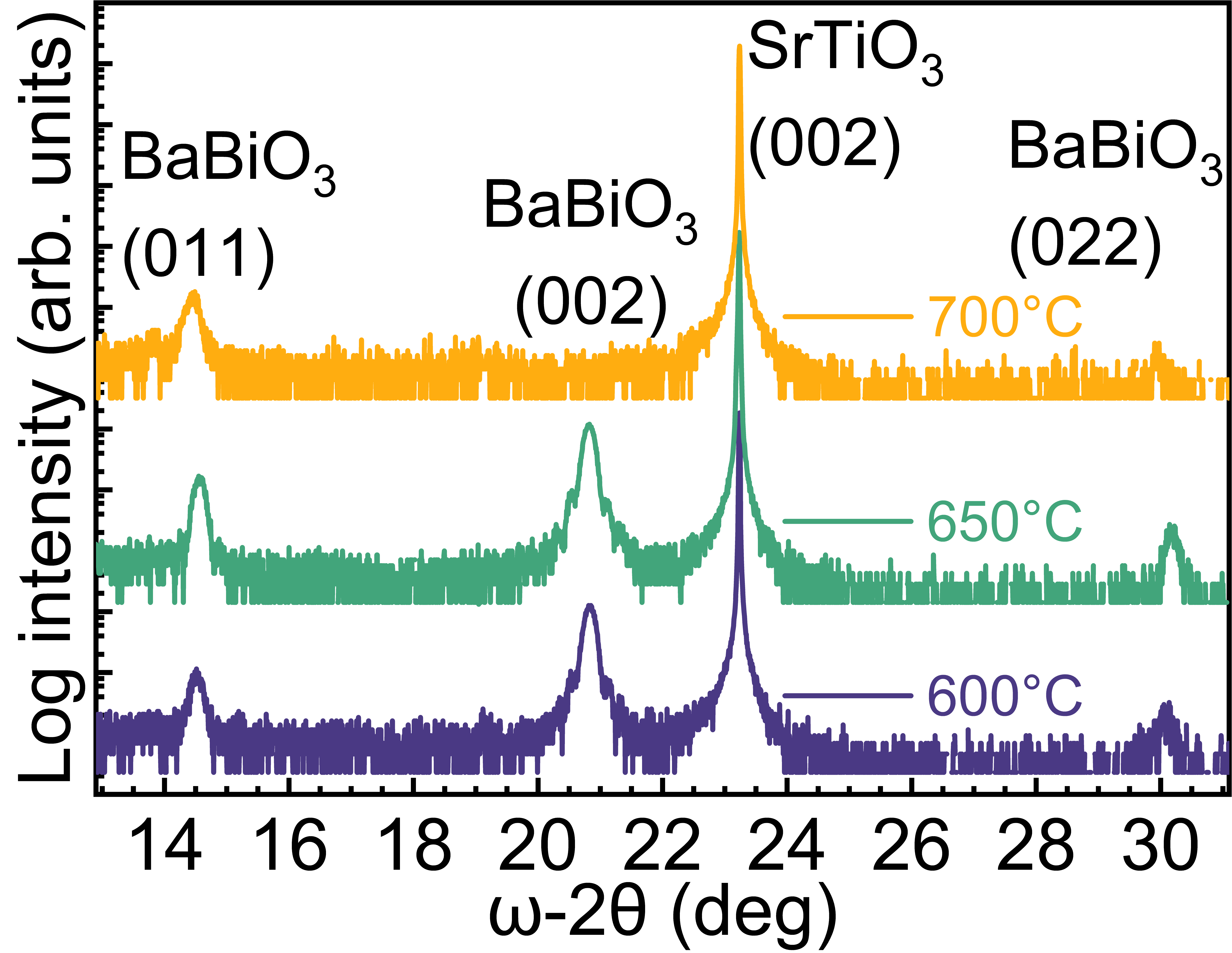}      
		
		\put(0, 79){(a)}      
		
		\put(101, 0){\includegraphics[width=0.245\textwidth]{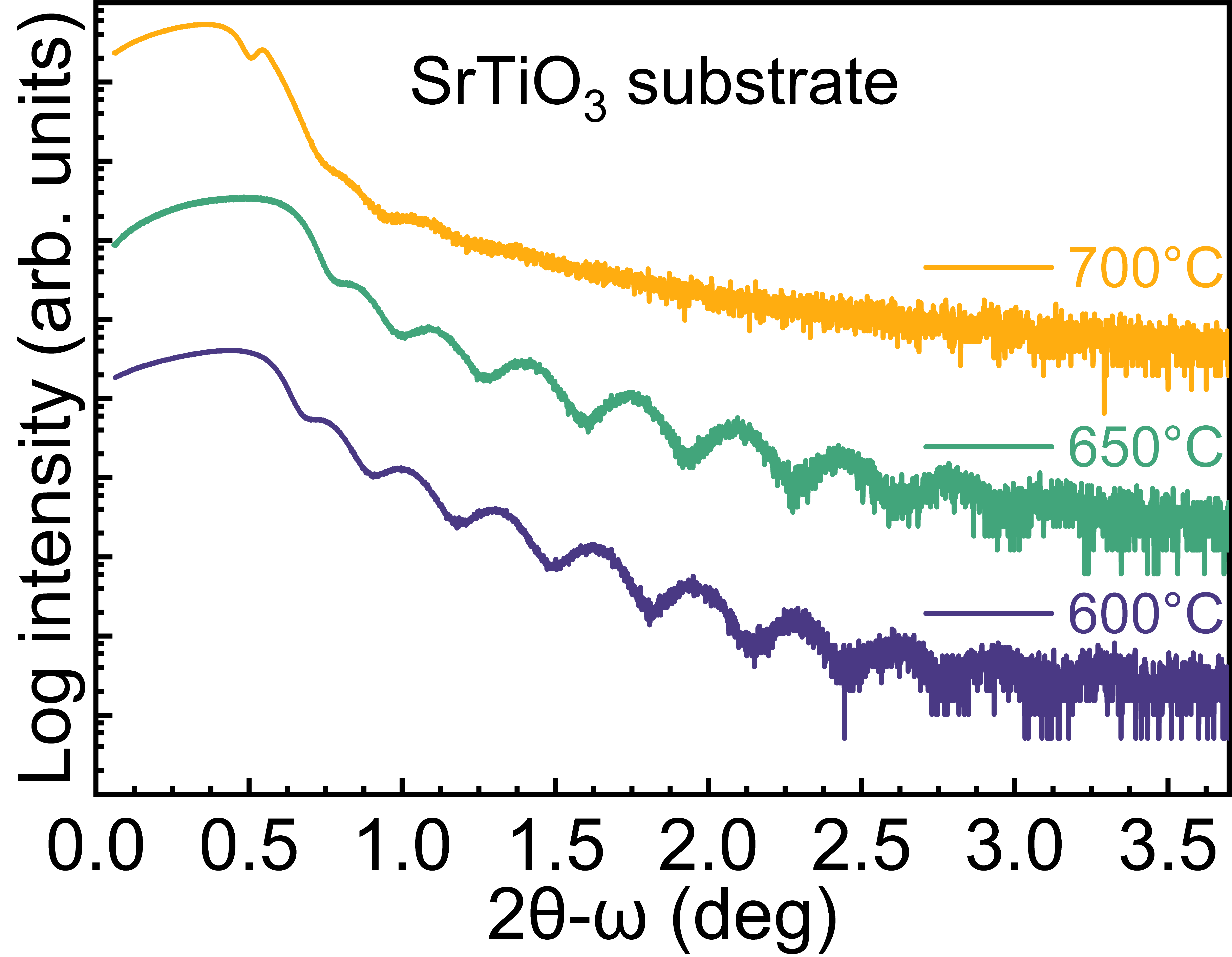}}      
		
		\put(101, 79){(b)}      
		
		\put(0, -60){\includegraphics[width=0.49\textwidth]{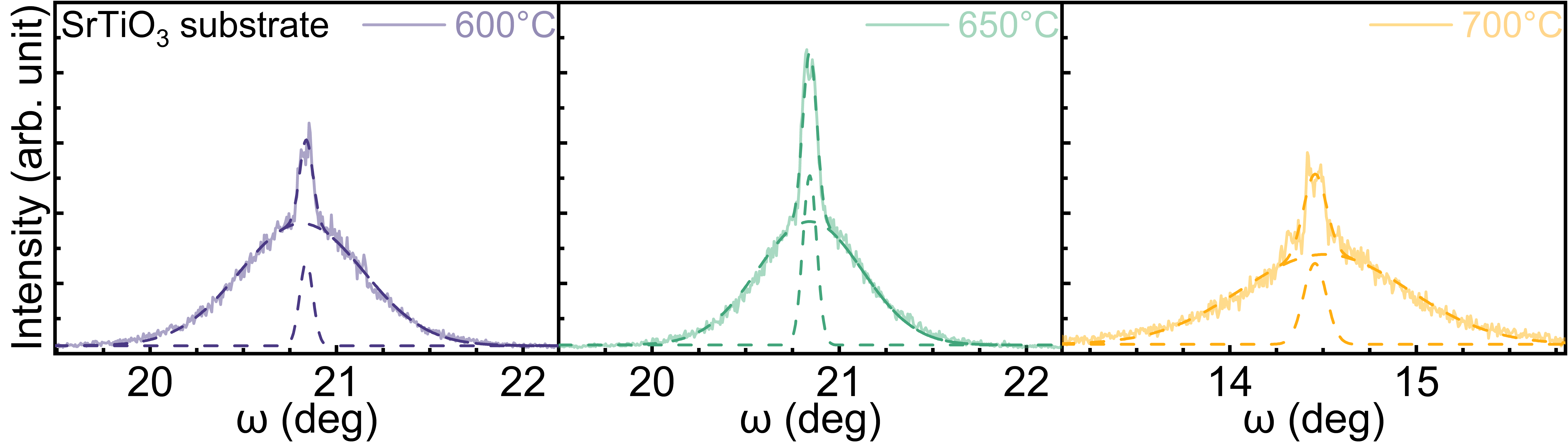}}      
		
		\put(0, -2){(c)}      
		
		\put(20, -103){\includegraphics[width=0.1\textwidth]{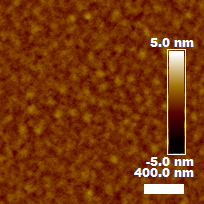}}      
		
		\put(83, -103){\includegraphics[width=0.1\textwidth]{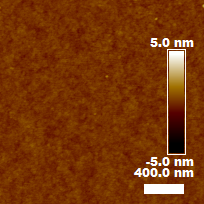}}      
		
		\put(147, -103){\includegraphics[width=0.1\textwidth]{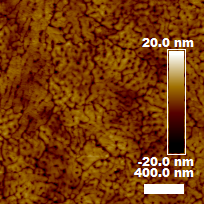}}     
		
		\put(10, -66){(d)}      
		
		\put(0, -210){\includegraphics[width=0.25\textwidth]{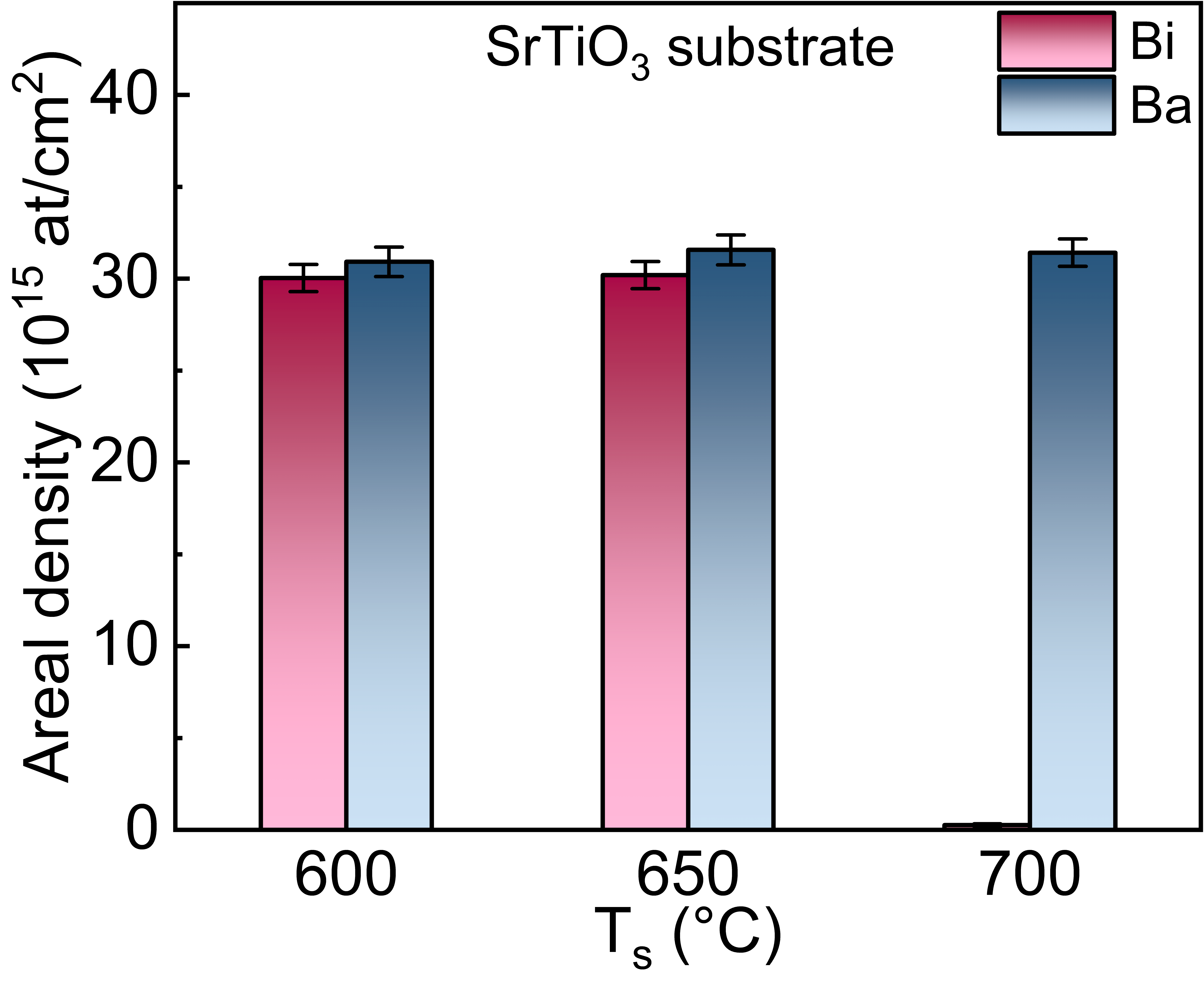}}      
		
		\put(0, -128){(e)}      
		
		\put(126, -215){\includegraphics[width=0.17\textwidth]{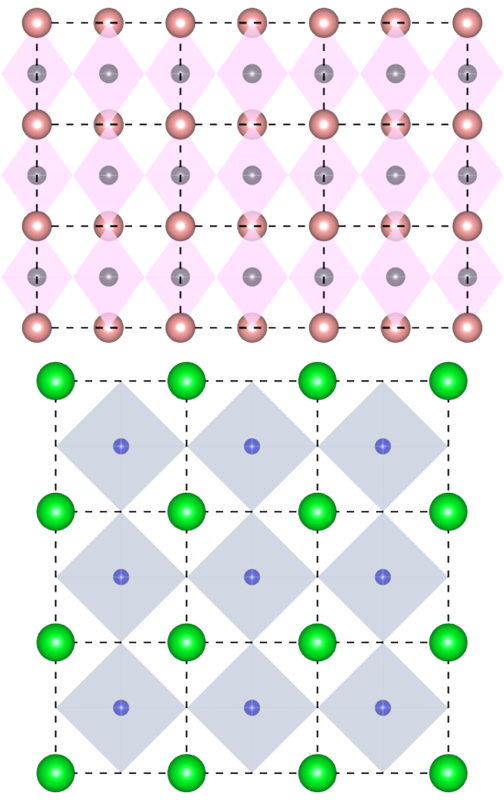}}      
		
		\put(98, -112){(f)}      
		
		\put(116.5, -193.5){  \begin{tikzpicture} 
				
				\draw[dashed] (0,0) -- (0.5,0); 
				
		\end{tikzpicture}} 
		
		\put(116.5, -175.5){  \begin{tikzpicture} 
				
				\draw[dashed] (0,0) -- (0.5,0); 
				
		\end{tikzpicture}} 
		
		\put(106, -187){3.905 \AA}      
		
		\put(111, -210){  \begin{tikzpicture} 
				
				\draw[->, thick] (0,0) -- (0.3,0); 
				
		\end{tikzpicture}}

		\put(111, -210){  \begin{tikzpicture} 
				
				\draw[->, thick] (0,0) -- (0,0.3); 
				
		\end{tikzpicture}}

		\put(111, -218){[100]}      
		
		\put(95, -210){[010]}      
		
		\put(118, -153){  \begin{tikzpicture} 
				
				\draw[dashed] (0,0.05) -- (1.5,-1); 
				
		\end{tikzpicture}} 
		
		\put(118, -145.5){  \begin{tikzpicture} 
				
				\draw[dashed] (0,0.05) -- (1.5,-1); 
				
		\end{tikzpicture}}

		\put(100, -120){3.78 \AA}      
		
	\end{overpic}     
	
	\vspace{9.8cm}     
	
	\caption{Effect of varying T$_s$ (600$^{\circ}$C, 650$^{\circ}$C, and 700$^{\circ}$C) on growing BBO on STO(001) substrate based on different characterization techniques is shown. (a) XRD, (b) XRR, (c) RC, (d) AFM, and (e) RBS data. (f) Schematic shows crystal structure of substrate and layer. Green balls: Sr ions, blue: Ti ions, orange: Ba ions, purple: Bi ions. Planes facing the reader are STO(001) and BBO(011). Dashed lines are for illustration of the epitaxial relationship: BBO$_{\text(011)}$[110]$\parallel$STO$_{\text(001)}$[100].}    
	
	\label{fig1}     
	
\end{figure}
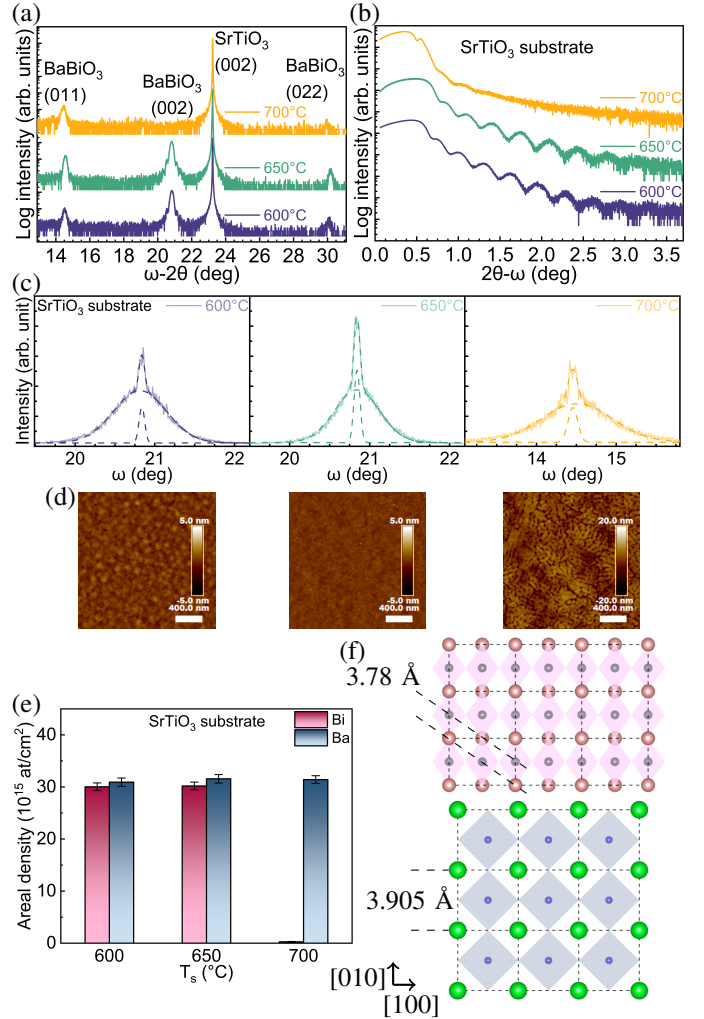

 In Fig. \ref{fig1}, variation in T$_s$ is investigated for thin films grown at (600$^{\circ}$C, 650$^{\circ}$C, and 700$^{\circ}$C) on STO substrate. Metallic flux was kept at J$_\text{Bi}$ = 7 A/s and J$_\text{Ba}$ = 1.7 A/s. According to the symmetric out-of-plane XRD scans in Fig. \ref{fig1}(a), diffraction peaks roughly at 20.8$^{\circ}$ and 14.5$^{\circ}$ are observed for thin films grown at 600$^{\circ}$C and 650$^{\circ}$C. This denotes that co-growth of BBO(001) \& BBO(011) is taking place. At 700$^{\circ}$C, only one diffraction peak at is observed 14.54$^{\circ}$. Crystalline quality can be evaluated based on the RC scans in Fig. \ref{fig1}(c), which show a broad peak corresponding to interfacial area with high defects and dislocation density in addition to a sharp peak indicating a highly ordered epitaxial areas in the film \cite{tellekamp2016molecular}. For 600$^{\circ}$C \& 650$^{\circ}$C, the sharp peaks have a full width at half maximum (FWHM) of 0.084$^{\circ}$ \& 0.088$^{\circ}$, respectively, indicating high quality epitaxy at these temperatures. XRR data in Fig. \ref{fig1}(b) for thin films grown at 600$^{\circ}$C and 650$^{\circ}$C show that surface and interface roughness are low with clearly distinguishable Kiessig fringes, and thickness of 30 nm and 27 nm, respectively. On the other hand, thin film grown at 700$^{\circ}$C is of higher roughness and thickness of 35 nm, extracted with lower accuracy due to the roughness induced oscillation rapid damping \cite{yasaka2010x}. AFM's surface texture of the film grown at 700$^{\circ}$C is visibly different in comparison with the other two films, with root mean square (RMS) roughness of 3.48 nm, while it is below 0.4 nm for the lower temperatures, as illustrated in Fig. \ref{fig1}(d). RBS data in Fig. \ref{fig1}(e) shows that films grown at 600$^{\circ}$C \& 650$^{\circ}$C are nearly stoichiometric; Bi/Ba = 0.97 \& 0.96, respectively, however, almost no Bi incorporation at 700$^{\circ}$C.

\begin{figure}[h]      
	
	\begin{overpic}[width=0.23\textwidth]{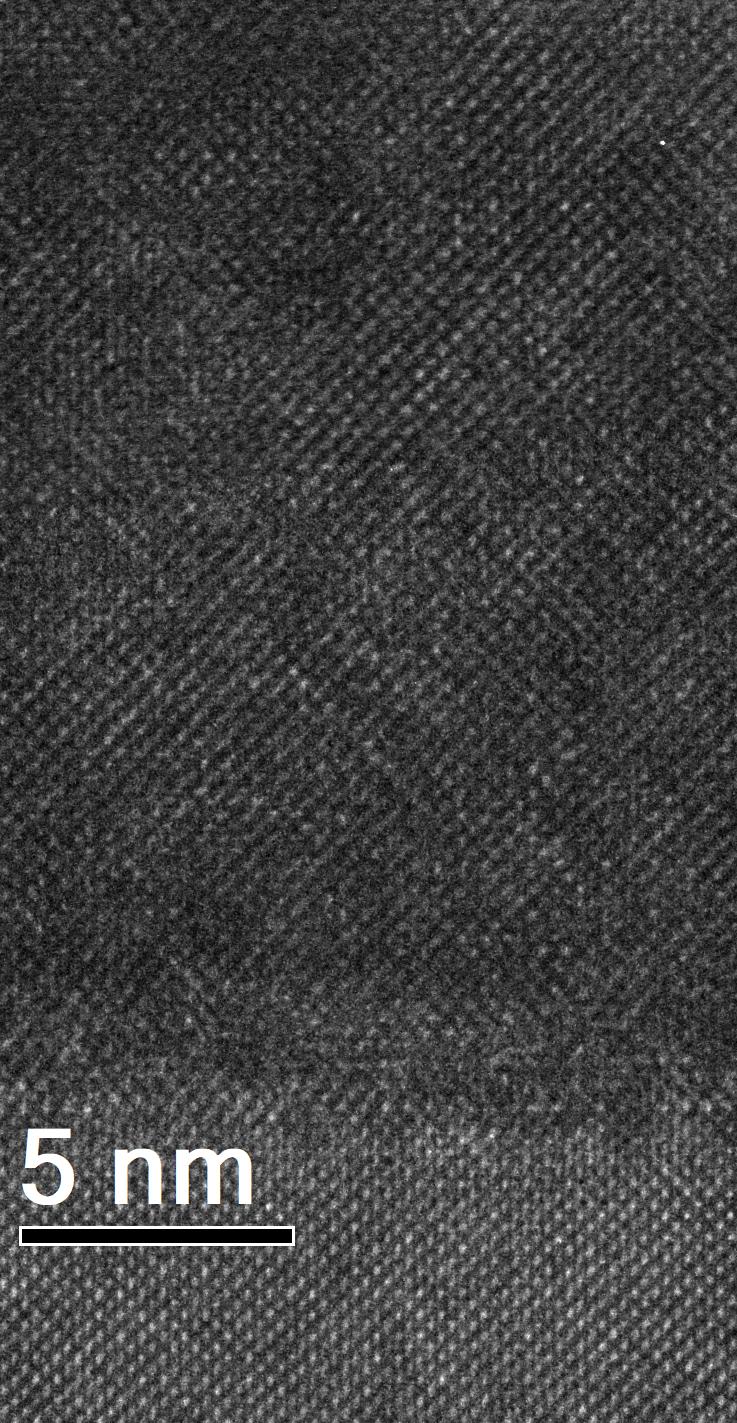}      
		
		\put(2, 96){\textcolor{white}{(a)}}      
		
		\put(54, 52){\includegraphics[width=0.245\textwidth]{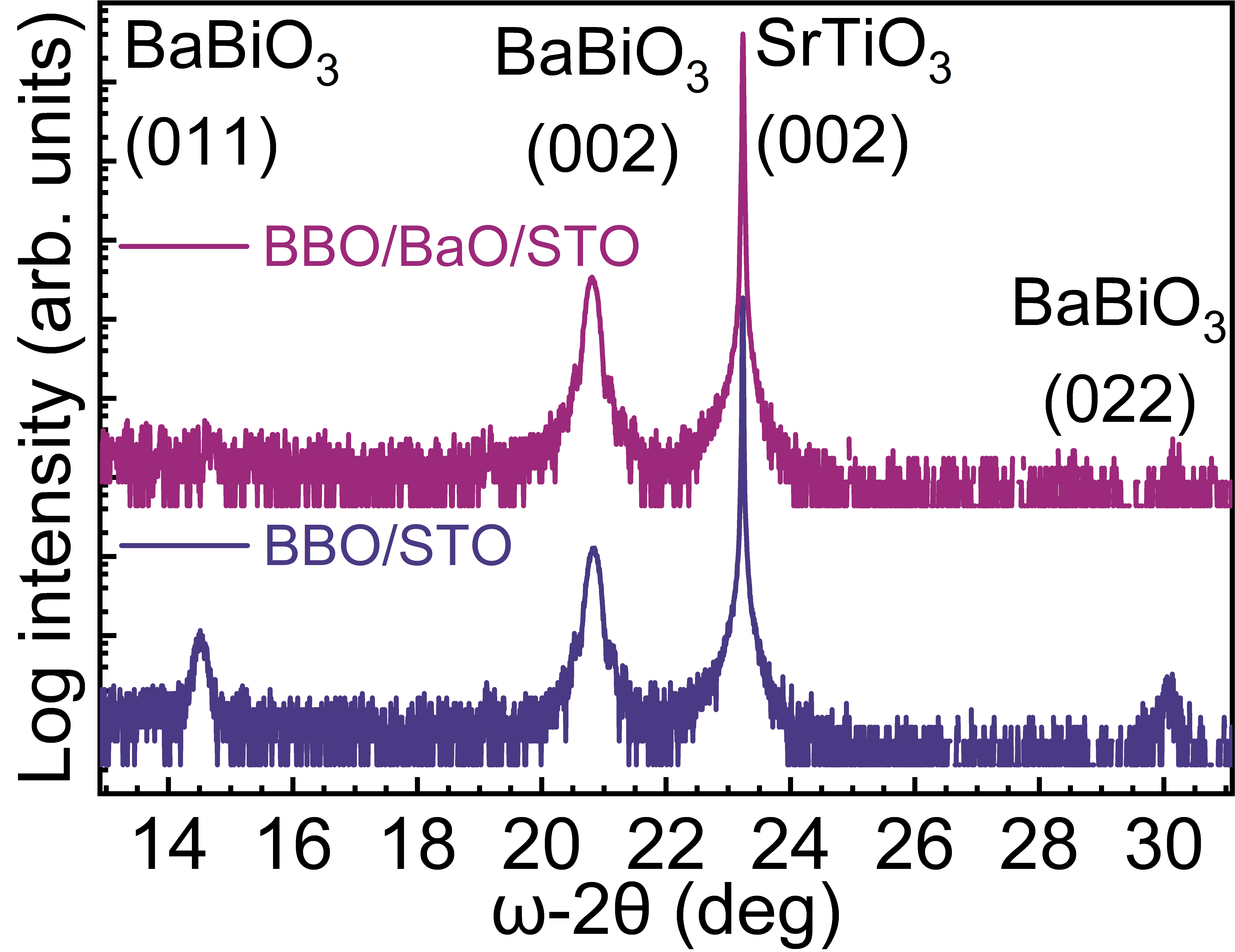}}      
		
		\put(54, 96){(b)}      
		
		\put(32,35){\textcolor{white}{{\large BaBiO$_{3}$}}}      
		
		\put(32,12){\textcolor{white}{{\large SrTiO$_{3}$}}}      
		
		\put(63, -8){\includegraphics[width=0.18\textwidth]{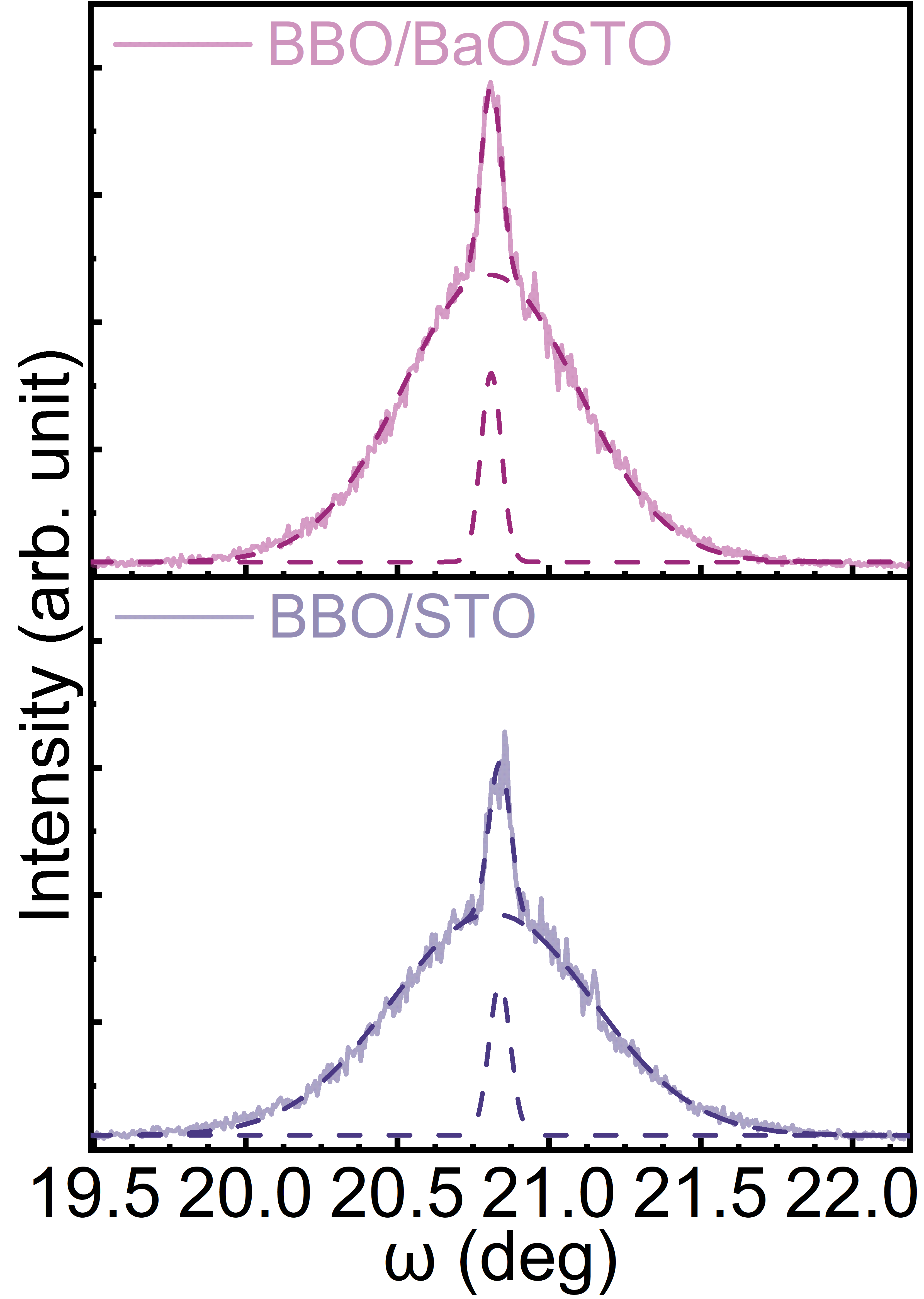}}      
		
		\put(62, 50){(c)}      
		
	\end{overpic}      
	
	\vspace{0.6cm}      
	
	\caption{Effect of implementing a BaO buffer layer on the orientation of BBO thin film grown on STO(001) substrate (600$^{\circ}$C, J$_\text{Bi}$ = 7 \AA/s) is shown. (a) TEM, (b) XRD, and (c) RC results.}      
	
	\label{fig4}      
	
\end{figure}

 One of the common approaches to stabilize a certain orientation in epitaxy is the use of a buffer layer \cite{zhang2009control, hsu2017controlled}. According to figure \ref{fig4}(b), it is clear that growing a BaO buffer layer helps in stabilizing only the (001) phase of BBO, with only one diffraction peak at $\omega_\text{BBO}$ = 20.81$^{\circ}$. Thin film with BaO buffer layer has an RC's FWHM of 0.079$^{\circ}$, in accordance with its sharp peak data in Fig. \ref{fig4}(c), denoting high crystalline quality. Growth of BBO(001) can also be observed from the TEM image in Fig. \ref{fig4}(a) with the aid of a BaO buffer layer. The reason why a monolayer of BaO as a buffer layer helps in stabilizing only the BBO(001) orientation is discussed in the discussions section \ref{discsec}.

\begin{figure}[htbp]      
	
	\begin{overpic}[width=0.245\textwidth]{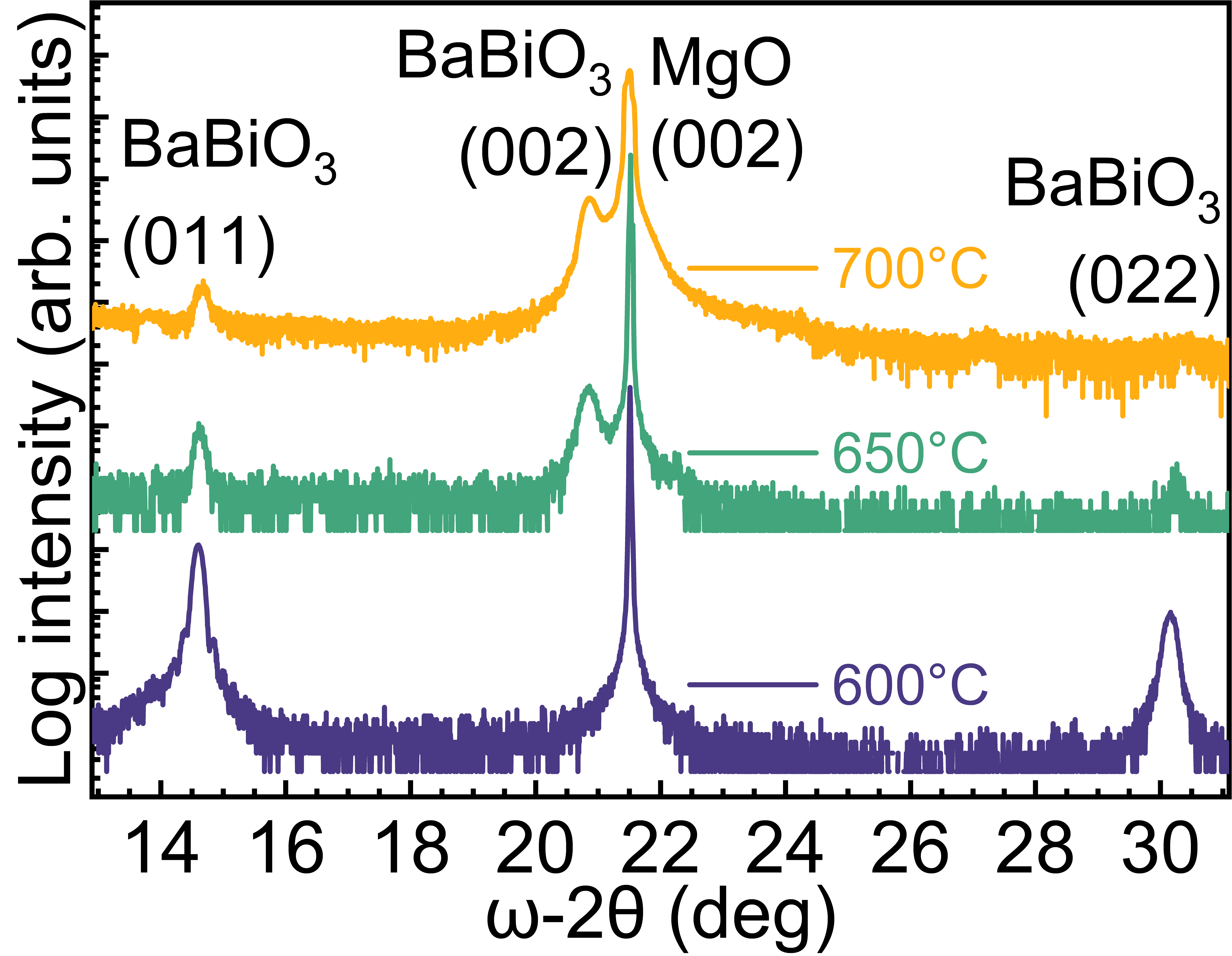}      
		
		\put(0, 79){(a)}      
		
		\put(101, 0){\includegraphics[width=0.245\textwidth]{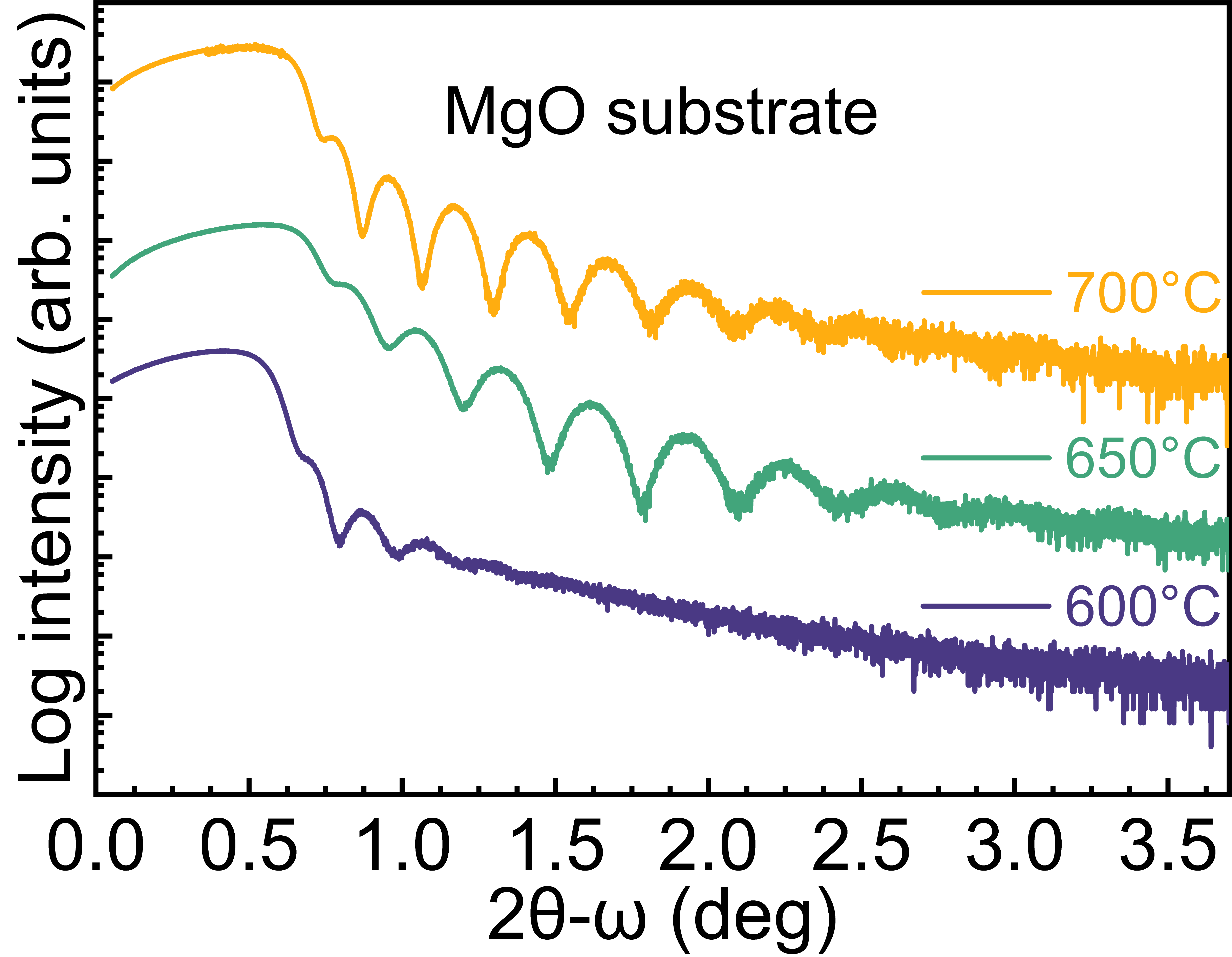}}      
		
		\put(101, 79){(b)}      
		
		\put(0, -60){\includegraphics[width=0.49\textwidth]{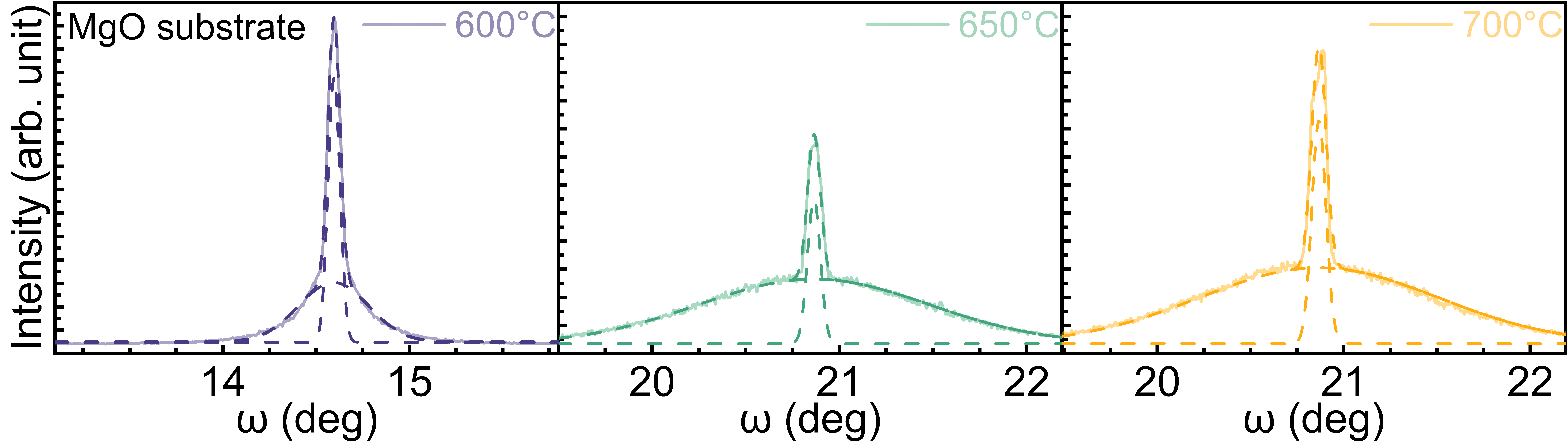}}     
		
		\put(0, -2){(c)}      
		
		\put(20, -103){\includegraphics[width=0.1\textwidth]{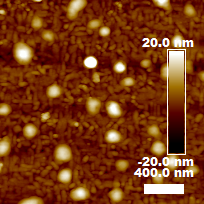}}      
		
		\put(83, -103){\includegraphics[width=0.1\textwidth]{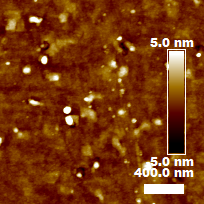}}      
		
		\put(147, -103){\includegraphics[width=0.1\textwidth]{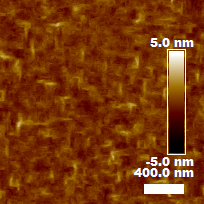}}      
		
		\put(10, -66){(d)}     
		
		\put(0, -210){\includegraphics[width=0.25\textwidth]{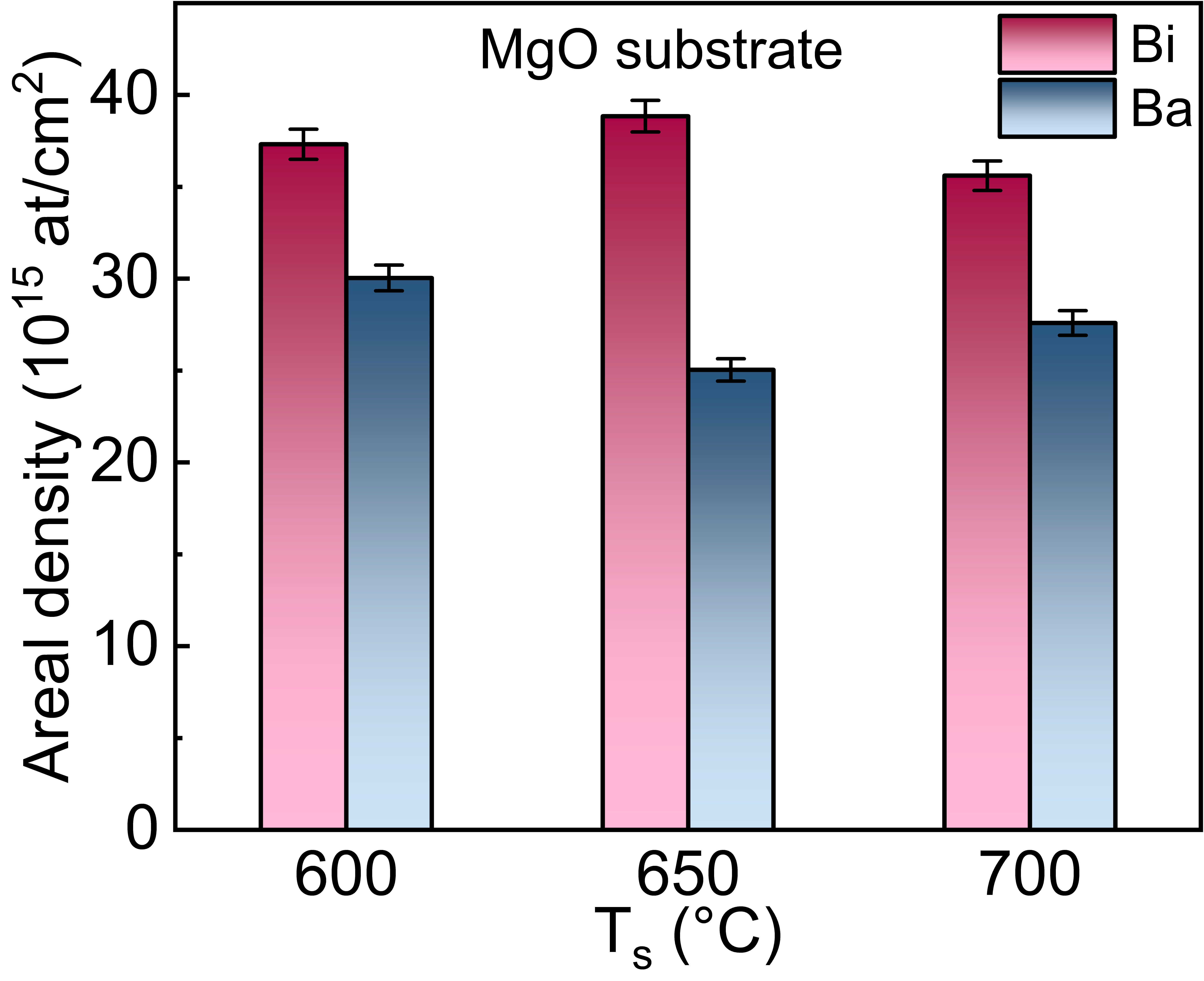}}     
		
		\put(0, -128){(e)}      
		
		\put(126, -219){\includegraphics[width=0.17\textwidth]{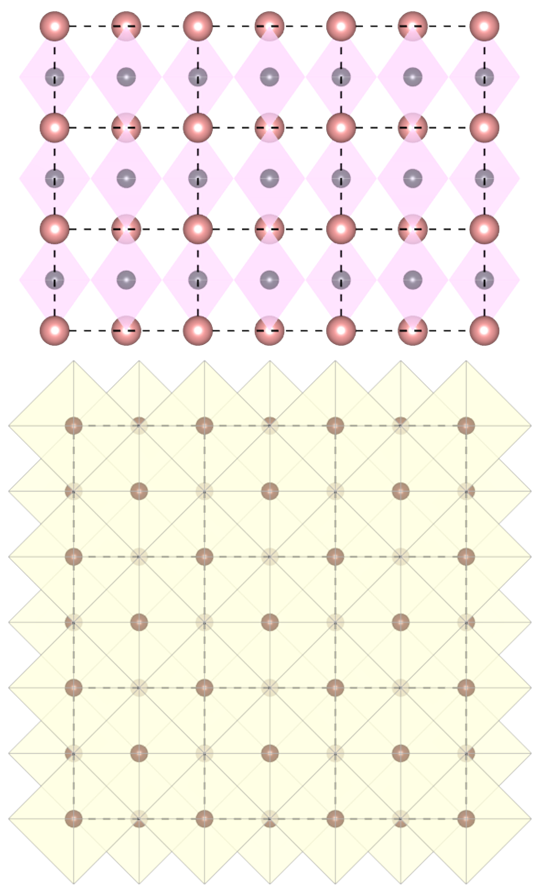}}      
		
		\put(98, -112){(f)}      
		
		\put(116.5, -193.1){  \begin{tikzpicture} 
				
				\draw[dashed] (0,0) -- (0.5,0); 
				
		\end{tikzpicture}} 
		
		\put(116.5, -176.3){  \begin{tikzpicture} 
				
				\draw[dashed] (0,0) -- (0.5,0); 
				
		\end{tikzpicture}} 
		
		\put(106, -187){4.212 \AA}      
		
		\put(111, -210){  \begin{tikzpicture} 
				
				\draw[->, thick] (0,0) -- (0.3,0); 
				
		\end{tikzpicture}} 
		
		\put(111, -210){  \begin{tikzpicture} 
				
				\draw[->, thick] (0,0) -- (0,0.3); 
				
		\end{tikzpicture}} 
		
		\put(111, -218){[100]}      
		
		\put(95, -210){[010]}      
		
		\put(120, -151){  \begin{tikzpicture} 
				
				\draw[dashed] (0,0.06) -- (1.5,-1); 
				
		\end{tikzpicture}} 
		
		\put(120, -144){  \begin{tikzpicture} 
				
				\draw[dashed] (0,0.05) -- (1.5,-1); 
				
		\end{tikzpicture}} 
		
		\put(100, -124){3.78 \AA}      
		
	\end{overpic}     
	
	\vspace{9.8cm}     
	
	\caption{Effect of varying T$_s$ (600$^{\circ}$C, 650$^{\circ}$C, and 700$^{\circ}$C) on growing BBO on MgO(001) substrate based on different characterization techniques is shown. (a) XRD, (b) XRR, (c) RC, (d) AFM, and (e) RBS data. (f) Schematic shows crystal structure of substrate and layer. Brown balls: Mg ions, orange: Ba ions, purple: Bi ions. Planes facing the reader are MgO(001) and BBO(011). Dashed lines are for illustration of the epitaxial relationship: BBO$_{\text(011)}$[110]$\parallel$MgO$_{\text(001)}$[100].}    
	
	\label{fig2}     
	
\end{figure}

For MgO substrate, Fig. \ref{fig2}(a) manifests that diffraction peaks at 14.6$^{\circ}$, with varied intensity, is noticed at all temperatures, but 20.85$^{\circ}$ only at 650$^{\circ}$C \& 700$^{\circ}$C. FWHM of the sharp peaks of the RC scans are 0.076$^{\circ}$, 0.081$^{\circ}$, and 0.087$^{\circ}$ for 600$^{\circ}$C, 650$^{\circ}$C, and 700$^{\circ}$C, respectively. High roughness of layer grown at 600$^{\circ}$C is evident based on the XRR data in Fig. \ref{fig2} (b). Diffuse scattering results in damping of reflected radiation intensity due to incoherent interference. Unlike smooth layers grown on STO substrate, XRR oscillations of the thin films grown on MgO at 650$^{\circ}$C and 700$^{\circ}$C do not have a fixed fringe width, therefore calculated thickness is 32±3 nm and 38±5 nm, respectively. According to AFM data in Fig. \ref{fig2}(d), surface of thin films grown on MgO substrates is getting smoother as temperature increases with RMS roughness of 4.49 nm at 600$^{\circ}$C, 0.91 nm at 650$^{\circ}$C, and 0.52 nm at 700$^{\circ}$C. Islands of average diameter of 90 nm and heigh of 27 nm, which are observed for thin film grown at 600$^{\circ}$C, are the root cause for this high roughness. RBS data in Fig. \ref{fig2}(e) indicates that all layers are Bi-rich; Bi/Ba = 1.24, 1.55, and 1.29 for 600$^{\circ}$C, 650$^{\circ}$C, and 700$^{\circ}$C, respectively.

\begin{figure}[h]    
	
	\begin{overpic}[width=0.29\textwidth]{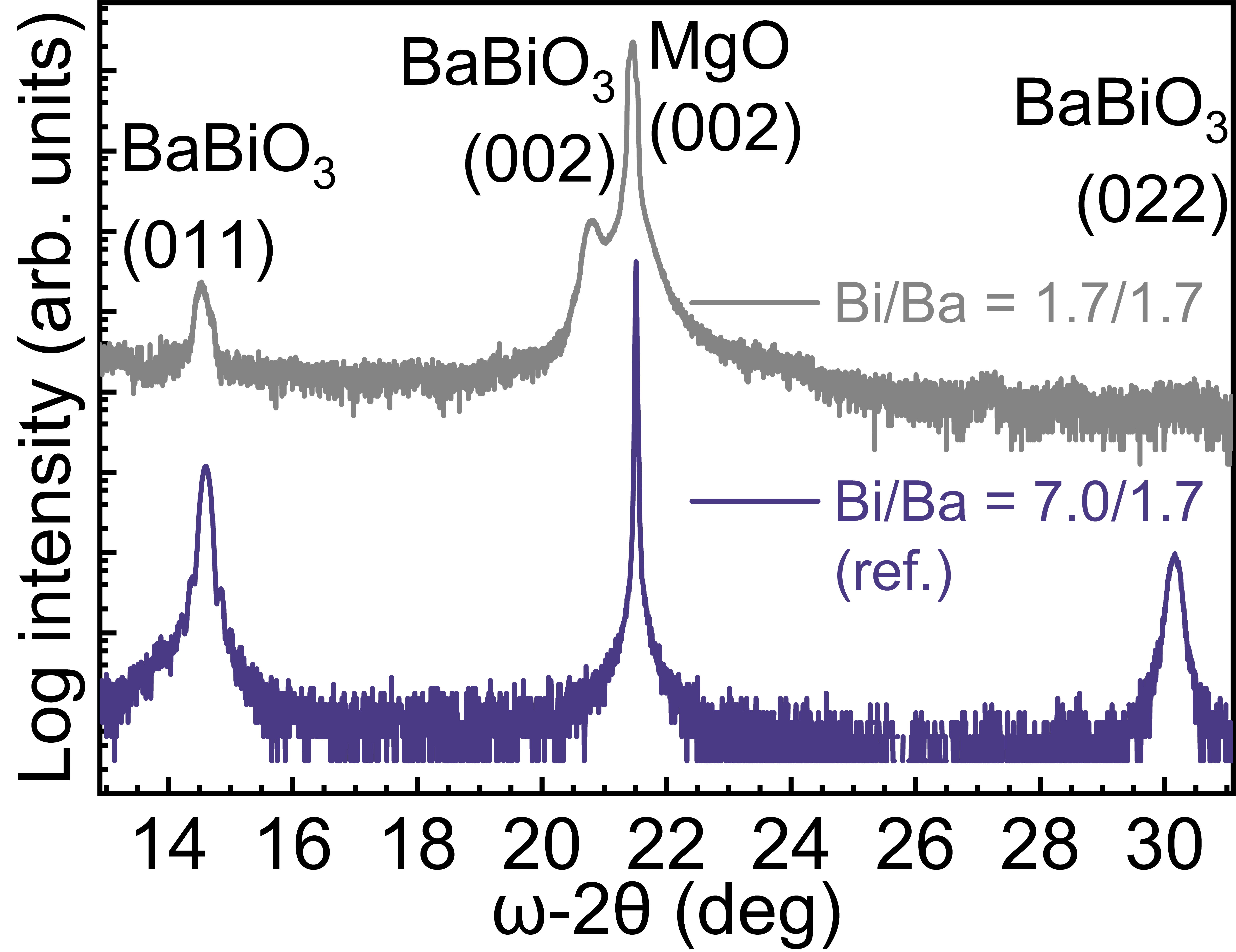}    
		
		\put(0, 78){(a)}    
		
		\put(101, 0){\includegraphics[width=0.2\textwidth]{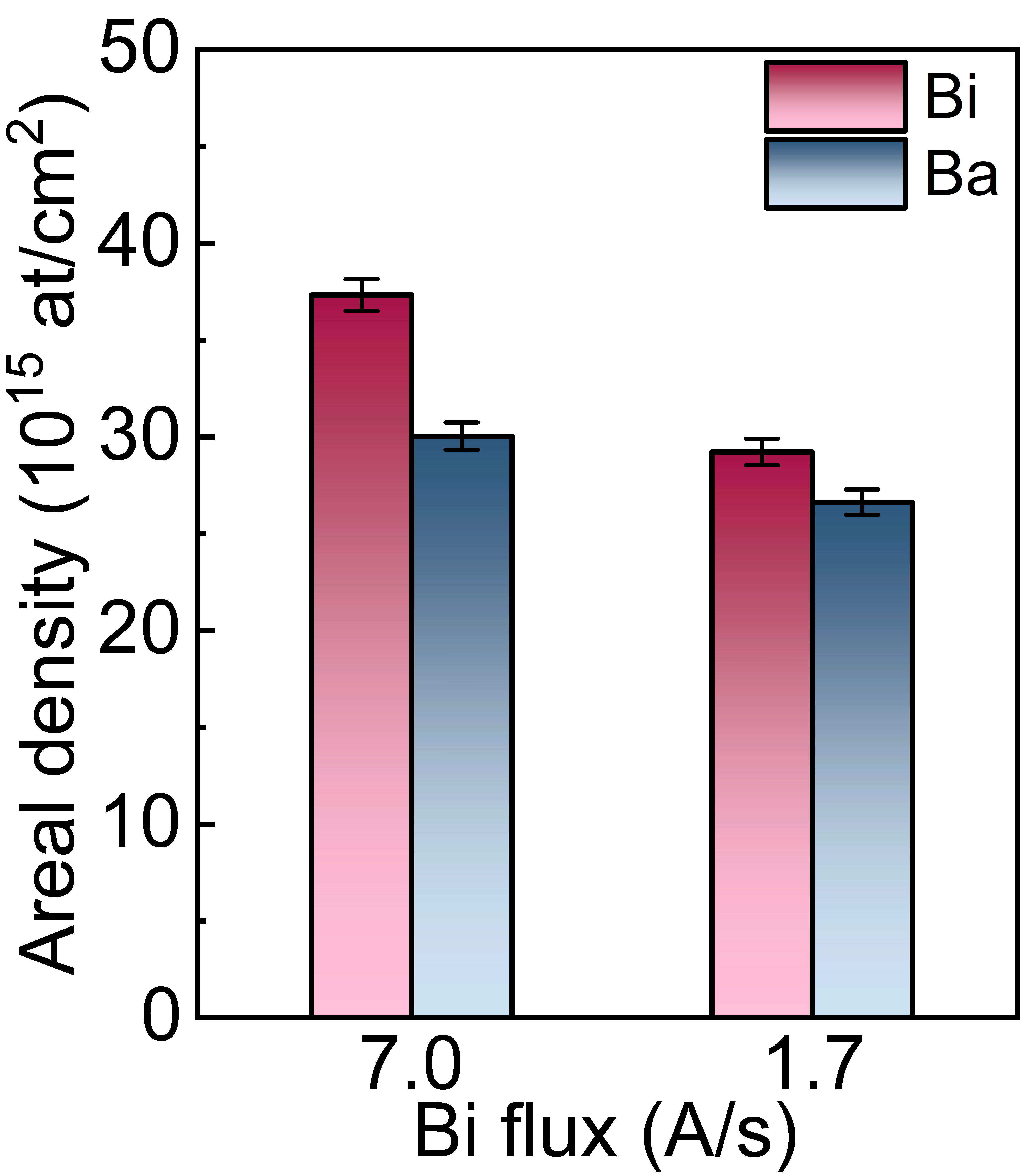}}    
		
		\put(100, 79){(b)}    
		
		\put(0, -74){\includegraphics[width=0.49\textwidth]{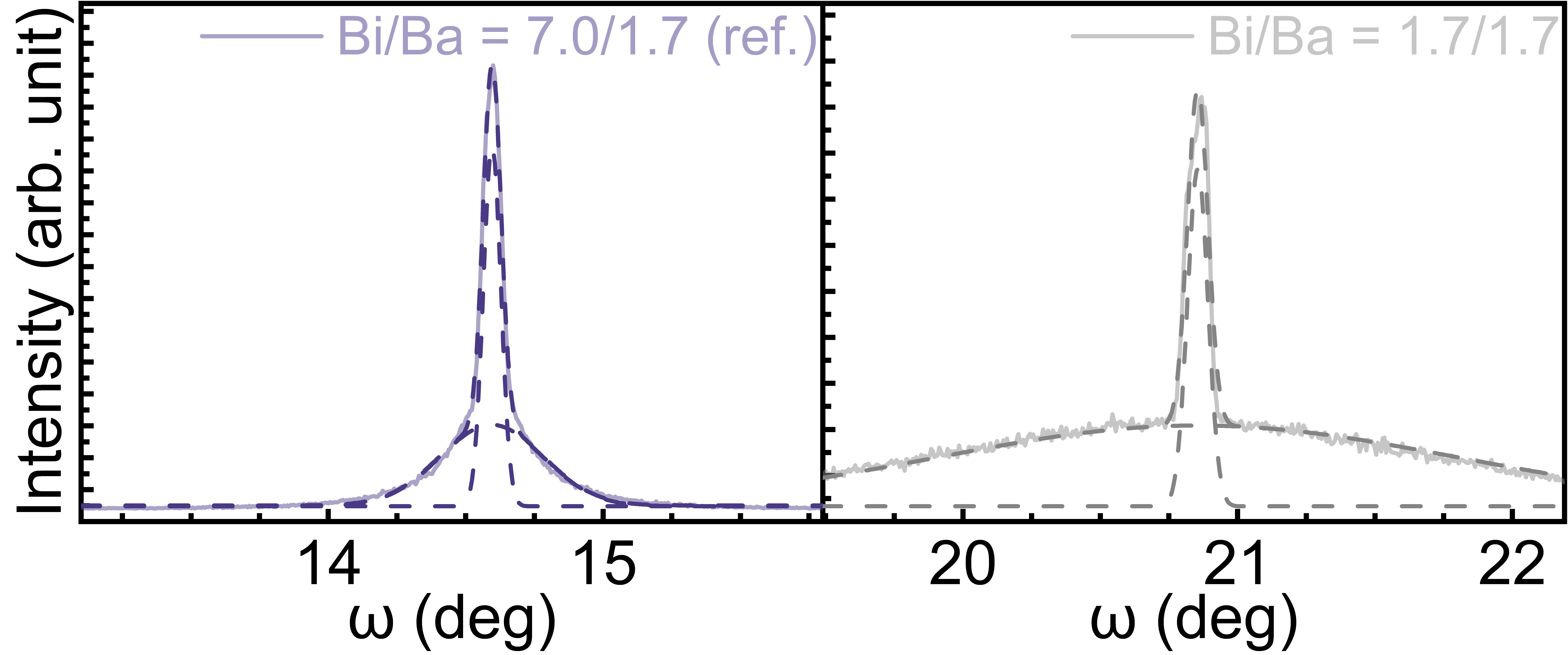}}    
		
		\put(0, -2.5){(c)}    
		
		\put(25, -129){\includegraphics[width=0.15\textwidth]{124}}    
		
		\put(25, -75){(d)}    
		
		\put(103, -129){\includegraphics[width=0.15\textwidth]{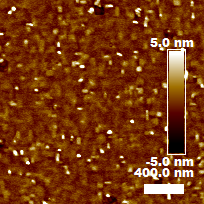}}    
		
	\end{overpic}    
	
	\vspace{6.9cm}    
	
	\caption{Effect of varying (Bi/Ba) flux ration, reducing it from 7/1.7 to 1.7/1.7 (while keeping T$_s$ at 600$^{\circ}$C), is illustrated for samples grown on MgO(001) substrates. (a) XRD symmetric scans, (b) RBS, (c) RC, and (d) AFM data.}    
	
	\label{fig3}    
	
\end{figure}

 Upon decreasing the flux ration from J$_\text{Bi}$/J$_\text{Ba}$ = 7.0/1.7 to J$_\text{Bi}$/J$_\text{Ba}$ = 1.7/1.7, a notable reduction in the diffraction peak at 14.57$^{\circ}$ and observation of another one at 20.81$^{\circ}$ can be seen from Fig. \ref{fig3}(a)'s out-of-plane XRD results. This result is associated with a more stoichiometric layer, based on RBS data in Fig. \ref{fig3}(b), with Bi/Ba of 1.09 (15\% reduction in film's Bi content). At J$_\text{Bi}$ of 1.7 \AA/s, BBO(001) can be grown with a sharp peak's FWHM of 0.088$^{\circ}$, according to RC data of Fig. \ref{fig3}(c). In addition, surface appears, in Fig. \ref{fig3}(d), to be smoother with RMS roughness of 0.98 nm, with no observation of large islands. The reported observations can be found in Tab. \ref{tab:my-table}.

\begin{table*}[t]
	\centering
		\renewcommand{\arraystretch}{1.5} 
	\setlength{\tabcolsep}{3pt} 
		\begin{tabular}{cccccccc}
			\multicolumn{1}{l}{} & MBE Growth condition & \multicolumn{2}{l}{$\omega$-measured ($^{\circ}$)} & BBO(011)/BBO(001)+BBO(011) & Thickness (nm) & RMS roughness (nm) & Bi/Ba (RBS data) \\
			\cline{2-8}
			\multirow{4}{*}{\rotatebox[origin=c]{90}{STO substrate}} 
			& 600 $^{\circ}$C                         & 14.41 & 20.81 & 7.56$\%$   & 30     & 0.39 & 0.97 ± 0.03 \\
			& 650 $^{\circ}$C                         & 14.56 & 20.80 & 11.81$\%$  & 27     & 0.23 & 0.96 ± 0.03 \\
			& 700 $^{\circ}$C                         & 14.45 & --    & --         & 35     & 3.48 & 0.008 ± 0.002 \\
			& 600 $^{\circ}$C (with BaO buffer layer) & 14.41   & 20.81 & 1.62$\%$   & 30    &    &  \\
			\hline
			\multirow{4}{*}{\rotatebox[origin=c]{90}{MgO substrate}} 
			& 600 $^{\circ}$C                         & 14.60 & --    & 100$\%$    & 33±4   & 4.49 & 1.24±0.04 \\
			& 650 $^{\circ}$C                         & 14.60 & 20.86 & 6.44$\%$   & 32±3   & 0.91 & 1.55±0.05 \\
			& 700 $^{\circ}$C                         & 14.68 & 20.87 & 2.37$\%$   & 38±5   & 0.52 & 1.29±0.04 \\
			& 600 $^{\circ}$C (J$_{\text{Bi}}$/J$_{\text{Ba}}$ = 1.7/1.7)       & 14.57 & 20.81 & 54.49$\%$  &  25  & 0.98 & 1.09 ± 0.04 \\
			\hline
		\end{tabular}%
	\caption{Summary table for the reported results. Flux ratio of J$_{\text{Bi}}$/J$_{\text{Ba}}$ = 7.0/1.7 are used, except if stated elsewise. Results include temperature variation experiments for both substrates as well as the use of BaO buffer layer on STO substrate and reducing J$_{\text{Bi}}$ on MgO substrate. All layers are grown at plasma conditions of 600 W, then cooled down to room temperature at the same plasma conditions with a rate of 10$^{\circ}$C/minute. (--) denotes "not observed", while (blank) denotes "not measured". XRD peaks are fitted using Fityk program to be able to quantify the orientation competition \cite{wojdyr2010fityk}.}
	\label{tab:my-table}
\end{table*}

\section{Discussions}
\label{discsec}

 Either the growth of BBO(011) or co-growth of BBO(011) and BBO(001) are observed growing the material on either STO or MgO substrates, according to XRD data. BBO(011) has been resulted before as the preferred orientation, either solely or in combination with BBO(001), when grown on STO(001) substrate by MBE \cite{iyori1995preparation}. This is contradicting its growth by pulsed laser deposition (PLD), where BBO(001) was obtained on STO(001), with a spontaneously occurring structurally different reconstruction layer of almost 1 nm at the interface \cite{zapf2018domain, bouwmeester2019stabilization, jin2020atomic}.  BBO(011) was previously also obtained on MgO(001) substrate despite the low lattice mismatch \cite{hellman1989molecular} of around 3\% compared to that with STO of around 11\%. This raises a question on how two in-plane domain orientation coexist in a similar way for the two different substrates. As it is odd to obtain (011) orientation on either STO(001) or MgO(001) (both with non-polar surfaces), especially that the BBO(011) orientation is polar. Additionally, surface energies of (011) orientations are higher than these of (001) orientations for most perovskite compounds, as studied by DFT, regardless of their terminations \cite{eglitis2023review, zhong2019first}. Therefore, same orientation epitaxy is expected for growth of bismuthate, under certain thermodynamic conditions.

 However, for explaining the growth of BBO(011) on STO substrate, the problem has to be treated from epitaxy point of view. At the interface, this orientation seems to have a relationship with the substrate of: BBO$_{\text(011)}$[110]$\parallel$STO$_{\text(001)}$[100], this way half of BBO(011)'s unit cell diagonal, with lattice spacing of 3.78 \AA, is aligning commensurately with one unit cell of STO, lowering the lattice mismatch to only -3 \%. This orientation, according to AFM results, are as smooth as BBO(001) orientation, unlike when it is grown on MgO substrate. This is due to the fact that with a large lattice mismatch of  -11 \%, for BBO$_{\text(011)}$[110]$\parallel$MgO$_{\text(001)}$[100], high strain energy drives the system into randomly forming 3D islands. 650$^{\circ}$C does not appear to be giving enough energy for the adatoms to find their lowest energy sites, which intuitively is found where BBO(001) is majorly formed when temperature is raised to 700$^{\circ}$C, with much smoother surface. It can be deduced that the presence of BBO(011) orientation for thin films grown on MgO substrate is associated with island growth and high surface roughness, as can be seen in Fig. \ref{figS1}. Atomic schematics showing the lattice plane BBO(011) are presented relative to STO and MgO substrates, in Fig. \ref{fig1}(f) and Fig. \ref{fig2}(f), respectively. Such orientation competition was observed multiple times in literature for oxide materials such as: STO on CeO$_2$ \cite{ye2021orientation}, Ba$_{0.8}$Sr$_{0.2}$TiO$_3$ on LaAlO$_3$ \cite{queralto2016disentangling}, and MgO on ZnO \cite{xiao2013competition}.

  BBO(001) was earlier grown on STO(001) substrate using PLD. A reported naturally occurring wetting layer at the interface is believed to uncouple the layer from the substrate, by dislocations formation and enabling domain matching epitaxy \cite{zapf2018domain, bouwmeester2019stabilization, jin2020atomic}. Detailed TEM study showed that this reconstructed layer is consisting of a double layer in the form of $\delta$-Bi$_\text{2}$O$_\text{3}$ with a fluorite structure and STO$_{\text(001)}$[100]$\parallel$$\delta$-Bi$_\text{2}$O$_\text{3}$$_{\text(001)}$[110]$\parallel$BBO$_{\text(001)}$[100] orientation\cite{jin2020atomic}. BaO buffer layer with in-plane lattice spacing of 5.53 \AA\ is grown in registry with STO with epitaxial relationships of BaO[100]$\parallel$STO[110], where a$_{\text{STO}}\sqrt{2}$ = 5.52 \AA. From epitaxy point of view, this facilitates the formation of the fluorite $\delta$-Bi$_\text{2}$O$_\text{3}$ with lattice constant of a$_{\delta\text{-Bi$_{\text{2}}$O$_{\text{3}}$}}$  = 5.65 \AA. \cite{jin2020atomic}, which ultimately enables a single orientation growth of BBO(001) on STO(001) substrate.

 Thin film grown on STO substrate at 700$^{\circ}$C has no bismuth incorporated due to the elemental low sticking coefficient as temperature increases. However, observation of a diffraction peak at 14.54$^{\circ}$ denotes the presence of an intact layer occurring with no bismuth incorporated, which could be BaO thin film. This claim is supported by surface image by AFM which demonstrates a cracked structure, possibly by absorbing moisture upon exposure to ambient forming of Ba(OH)$_2$, due to its hygroscopic nature \cite{jahoda1957fundamental}. Additionally, oscillation amplitude of XRR data is less pronounced for this layer compared to all others because of the less contrasted electron densities with respect to the underlying substrate ($n_\text{BaO}$ = 1.44E24 e/cm$^3$, $n_\text{BBO}$ = 1.58E24 e/cm$^3$, $n_\text{STO}$ = 1.32E24 e/cm$^3$).

 On the other hand, XRR's oscillation period for thin films grown on MgO is not fixed, which points out film's inhomogeneous composition. This is validated by RBS result of Bi-rich layers, even at high temperature. This is contradicting the case of growing BBO on STO substrates. In fact, BBO follows adsorption controlled regime on STO substrate, meaning that even at high flux of Bi, only a limited amount is incorporated within the growth windows, which satisfy self-regulating stoichiometric level \cite{hellman1990adsorption}. The pronounced decrease in the incorporated Bi by lowering its flux refers to the fact that growth of BBO on MgO substrate does not follow adsorption-controlled regime, and deposition is flux-limited. Indeed, kinetics of adatoms depends strongly on the substrate chemical termination and its available adsorption sites. It could be the case that in order to access adsorption-controlled regime of BBO on MgO substrate, growth temperature needs to be increased beyond the limit of the MBE system \cite{brahlek2018frontiers}.

\section{Conclusion}

 In this paper, co-growth of BBO(001) \& BBO(011) is observed for both STO and MgO substrates. Studying the effects of substrate temperature, flux ratio, and interface engineering enables better understanding of the mechanism of how the double orientations are developed. Different pathways for controlling the orientation of the grown thin film are described depending on the underlying substrate. For STO, at 700$^{\circ}$C, no Bi is incorporated, and at 600$^{\circ}$C \& 650$^{\circ}$C doubly oriented BBO(001) \& BBO(011) are obtained. Growth of BBO(011) on STO is facilitated by the commesurate epitaxial relationship: BBO$_{\text(011)}$[110]$\parallel$STO$_{\text(001)}$[100]. In order to discourage the growth of BBO(011) and control the epitaxial relation, a buffer layer of BaO is used. Interface engineering enables the growth of a reconstructed interfacial layer of $\delta$-Bi$_\text{2}$O$_\text{3}$, which allows for the domain matching epitaxy of BBO(001) on STO(001). For MgO, almost majorly oriented BBO(001) film is obtained only at 700$^{\circ}$C. Below this temperature, film is either doubly oriented or entirely BBO(011), with high surface roughness. Increasing substrate temperature helps in stabilizing BBO(001) orientation, which is expected with a small lattice mismatch of 3\%. Polar BBO(011)'s surface is expected to be less stable compared to that of BBO(001), therefore, it could be concluded that a balance between surface energies and strain energies determines the growth orientation. This defines a more structured way to control BBO crystal orientation, which is important for example when studying topological insulating behaviour, which is only foreseen for BBO(001) oriented crystals \cite{yan2013large}.

\section*{Acknowledgements}	

\quad The authors would like to thank hardware engineers Hans Costermans and Kevin Dubois for their dedicated support on the MBE cluster tool. This work has received funding from the European Research Council (ERC) under the European Union’s Horizon 2020 research and innovation program (grant agreement No 864483).





\section*{Supplementary material}

\renewcommand{\thefigure}{S\arabic{figure}}
\setcounter{figure}{0}

\begin{figure}[h]   
	
	\centering    
	
	\includegraphics[width=0.4\textwidth]{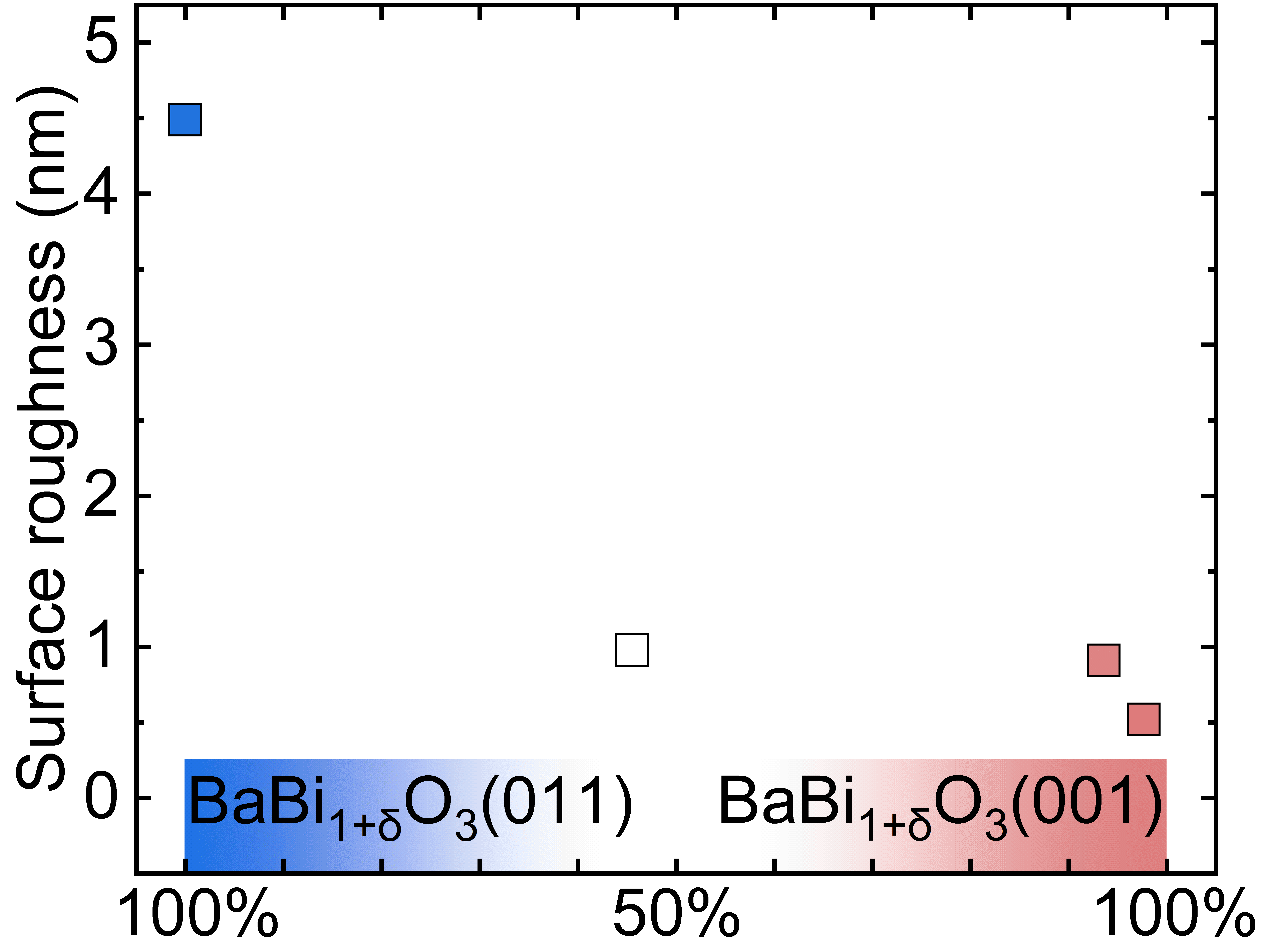}   
	
	\caption{Illustration shows the relationship between crystal orientation and surface roughness of BBO films grown on MgO substrate based on a summary of AFM data.}   
	
	\label{figS1}   
	
\end{figure}

\end{document}